\documentclass[aps,prl,twocolumn,superscriptaddress,showpacs,showkeys]{revtex4-1}
\usepackage[english]{babel}
\usepackage{amssymb}  
\usepackage{amsmath}
\usepackage{graphicx}
\usepackage{xcolor}
\usepackage[pdftex,colorlinks=true,linkcolor=blue,citecolor=blue,urlcolor=blue,filecolor=blue,breaklinks=true]{hyperref}  
\usepackage{url}
\definecolor{dark-blue}{rgb}{0.15,0.15,0.4}

\begin{document}

\title{Visualising the connection between edge states and the mobility edge in adiabatic and non-adiabatic topological charge transport}

\author{Mariya A.~Lizunova} 
\affiliation{Institute for Theoretical Physics, Utrecht University, Princetonplein 5, 3584 CC Utrecht, The Netherlands}
\affiliation{Institute for Theoretical Physics Amsterdam, University of Amsterdam, Science Park 904, 1098 XH Amsterdam, The Netherlands}
\author{Florian Schreck}
\affiliation{Institute of Physics, University of Amsterdam, Science Park 904, 1098 XH Amsterdam, The Netherlands}
\author{Cristiane Morais Smith}
\affiliation{Institute for Theoretical Physics, Utrecht University, Princetonplein 5, 3584 CC Utrecht, The Netherlands}
\author{Jasper van Wezel}
\email{vanwezel@uva.nl}
\affiliation{Institute for Theoretical Physics Amsterdam, University of Amsterdam, Science Park 904, 1098 XH Amsterdam, The Netherlands}

\pacs{71.45.Lr,73.43.Cd,72.20.Ee,73.20.At}

\keywords{topological transport, charge-density wave, mobility edge}

\begin{abstract}
The ability to pump quantised amounts of charge is one of the hallmarks of topological materials. An archetypical example is Laughlin's gauge argument for transporting an integer number of electrons between the edges of a quantum Hall cylinder upon insertion of a magnetic flux quantum. This is mathematically equivalent to the equally famous suggestion of Thouless' that an integer number of electrons are pumped between two ends of a one-dimensional quantum wire upon sliding a charge-density wave over a single wave length. We use the correspondence between these descriptions to visualise the detailed dynamics of the electron flow during a single pumping cycle, which is difficult to do directly in the quantum Hall setup, because of the gauge freedom inherent to its description. We find a close correspondence between topological edge states and the mobility edges in charge-density wave, quantum Hall, and other topological systems. We illustrate this connection by describing an alternative, non-adiabatic mode of topological transport that displaces precisely the opposite amount of charge as compared to the adiabatic pump. We discuss possible experimental realisations in the context of ultracold atoms and photonic waveguide experiments.
\end{abstract}

\maketitle

\section{Introduction}
The experimental discovery of the integer quantum Hall effect (IQHE) in GaAs~\cite{topin0} revealed the first example of a two-dimensional ($2D$) material that has an insulating bulk, but metallic edge states. The IQHE state is now understood to be part of a much larger class of topological insulators~\cite{topinsym1,topinsym2,topinsym3,topinsym3a,topinsym5,topinsym6,topinsym7,juricic}, which can be labelled by topological invariants such as the Chern number or spin-Chern number~\cite{intro9}. Although these indices determine the number and types of states localised on the edges of a finite sample, they may be computed entirely from bulk quantities, and depend on the symmetries of the bulk Hamiltonians. Different topological classes emerge depending on whether time-reversal, particle-hole and chiral symmetries are present or not~\cite{topinsymmetry1,topinsymmetry2}, and may be further refined using lattice symmetries~\cite{jorrit1,jorrit2,po1,po2,bernevig}.

A correspondence between the bulk topological invariants and the dynamics of edge states in the IQHE has been put forward by Laughlin~\cite{intro3,Halperin,intro9,intro3b}. It shows that adiabatically threading a single Aharonov-Bohm flux quantum $ \phi_0=h/e$ through the interior of an IQHE cylinder~\cite{intro3,Halperin}, results in a quantized number of electrons moving from one edge of the cylinder to the opposite. The amount of charge transported is given precisely by the bulk Chern number, defined as an integral over the $2D$ Brillouin zone~\cite{thouless}. The resulting cyclic charge transfer may be regarded as a dynamical manifestation of the IQHE~\cite{intro11}, and has been observed in a Corbino geometry~\cite{Dolgopolov,Prinz}. More generally, it is an example of the type of quantized adiabatic particle transport, or topological charge pumping, first proposed by Thouless~\cite{intro9,thouless,niu,intro10}. Experimentally, topological pumps have been realized in cold-atom~\cite{intro12,intro13} and single-spin~\cite{intro14} systems, and attracted attention in both the adiabatic~\cite{intro4,intro5} and non-adiabatic~\cite{intro6, intro7} regimes.
%
\begin{figure*}[t]
\includegraphics[width=0.9\textwidth]{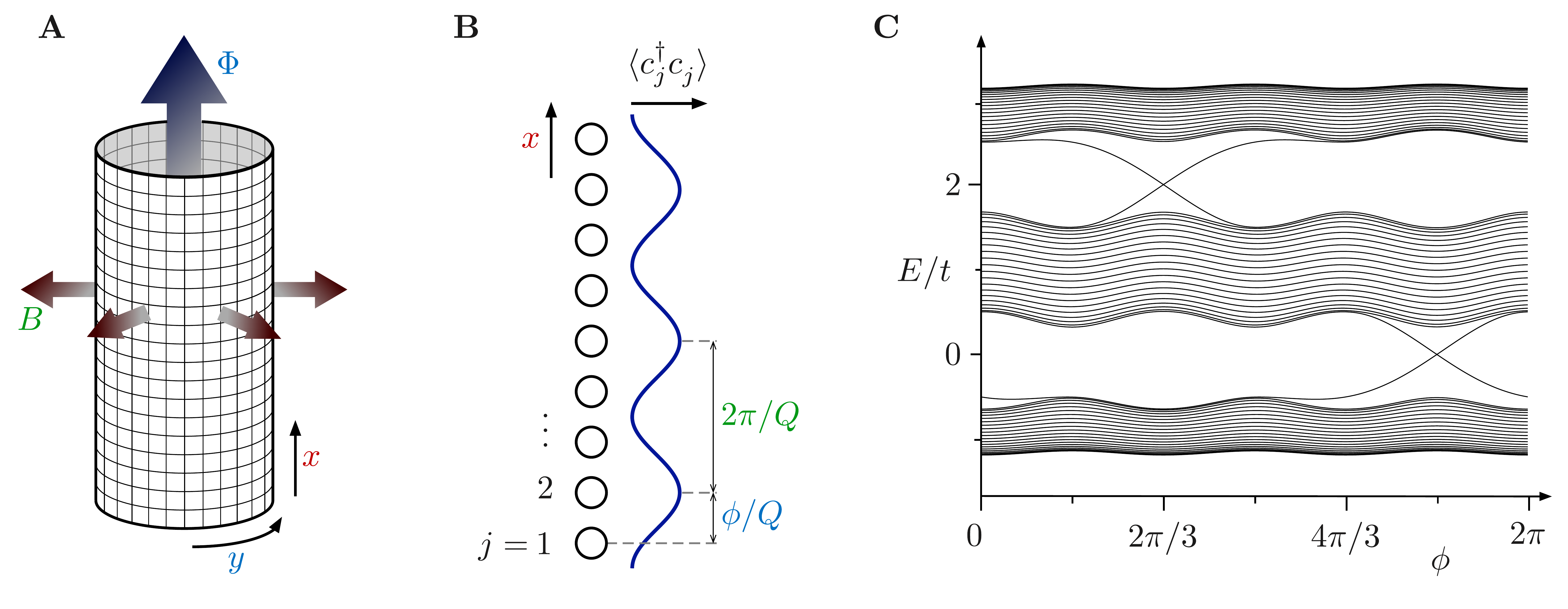}
\caption{\label{fig1} Schematic depictions of the quantum Hall cylinder used in Laughlin's gauge argument ({\bf A}), and the mean-field charge-density wave (CDW) used as a Thouless pump ({\bf B}). Both systems are described by Harper's equation, where the $x$-coordinate is a spatial direction in both cases. The perpendicular magnetic field $B$ in the cylinder plays the same role as the band filling $n$ or, equivalently, propagation vector $Q$ in the CDW. The phase of the charge-density modulation can be mapped to a gauge-independent combination of the $y$-component of momentum and the magnetic flux $\Phi$ piercing the quantum hall cylinder. {\bf C} The spectrum of the two systems is the same, including the presence of localised, in-gap modes at the edges of the $x$-direction. Notice that for the CDW, each value of $\phi$ represents a separate $1D$ system, while for the $2D$ quantum Hall cylinder $\phi$ corresponds to the $y$-component of momentum and states for all values of $k_y$ are simultaneously occupied.}
\end{figure*}

Despite its status as an archetype of topological transport, the detailed dynamics of exactly how electrons are transferred between edges as a quantum of flux is threaded through Laughlin's IQHE cylinder remains difficult to visualize. The straightforward comparison of electronic wave functions for values of the flux that differ by fractions of the flux quantum is hampered by the flux-dependence of the canonical momentum. In this paper, we circumvent this problem by using a well-known family of one-dimensional ($1D$) charge-ordered systems that can be mathematically mapped onto the $2D$ IQHE setup~\cite{cdw1}. This allows us to directly visualize the topological transport in real space and time, as electrons flow from one edge of the system to the other, and thus clarify the nature of the transport in both the charge-ordered and IQHE systems.

The rendering of edge state dynamics additionally reveals the in-gap states to be connected to the so-called mobility edge, which is the critical state separating extended from localized states in the bulk of a disordered system~\cite{RMP_me,est_me}. In the case of quantum Hall systems, the bulk extended states are known to be constricted to just a single energy in the centre of each (impurity-broadened) Landau level, as long as inter-Landau level scattering is negligible~\cite{RMP_me,Chalker_me}. Both edge states and the mobility edge are topological in nature~\cite{pruisken}, which we confirm by considering their common insensitivity to the presence of weak impurities. Together, the edge states and the mobility edges make up a single set of connected states winding around the electronic spectrum. We illustrate this connection of all the topological states by suggesting an alternative form of topological charge transport, based on a non-adiabatic variation of externally applied fields. The non-adiabatic charge pump transfers precisely the opposite amount of charge from its traditional adiabatic counterpart, and thus acts as an ``anti-Thouless'' pump. We suggest a possible experimental realisation in cold atomic gasses in an optical lattice.

\section{Mapping between $1D$ and $2D$}
Harper's equation, and more generally, a Mathieu equation~\cite{iqhecdw1,iqhecdw2,hofstadter}, can be interpreted both as a tight-binding model for the IQHE in $2D$, and as a family of mean-field descriptions for monatomic commensurate charge-density waves (CDW) in $1D$~\cite{cdw0,cdw1,Loss1,Loss2,Loss3}. The filling of the charge-ordered system, and hence its CDW periodicity, for example, can be taken to correspond to the magnetic field strength perpendicular to the surface of a quantum Hall cylinder, which determines the filling of its Landau levels~\cite{cdw1,Loss1,hofstadter}. Under the same mapping, the phase $\phi$ of the CDW order parameter then corresponds to a flux threading the quantum Hall cylinder~\cite{intro3}, while the spatial coordinate of the CDW chain is directly related to the spatial coordinate parallel to the axis of the $2D$ cylinder. The mapping is indicated schematically in figure~\ref{fig1}.

Upon varying the parameter $\phi$ adiabatically from $0$ to $2\pi$, the CDW slides along the $1D$ chain over precisely one wave length. The flux threading the IQHE system is increased by one quantum under the same variation. When introducing edges, we thus expect to find adiabatic transport of a quantised number of electrons from one side of the system to the other in both cases. In this way, the mapping relates the quantized adiabatic particle transport (Thouless pumping~\cite{thouless}) in charge-ordered systems to the topological transport between edge states upon insertion of a flux quantum in a quantum Hall cylinder (Laughlin's gauge argument~\cite{intro3}).

The bulk of both the CDW and IQHE systems is insulating, and only the edge states cross the Fermi level. By solving Harper's equation on a cylinder, the spectrum can be plotted as a function of momentum in the periodic direction. It consists of bulk bands and isolated topological edge states crossing the gaps between them. The edge states are protected, in the sense that the number of edge channels cannot be modified, as long as the bulk of the system remains gapped~\cite{intro3}. In a CDW system the periodic direction is given by the mean-field value of the phase variable $\phi$, as shown in figure~\ref{fig1}. Each value of the phase then corresponds to a single realisation of a CDW on the chain, which may or may not host edge states. The combined spectrum of the family of CDWs containing all values of $\phi$ coincides with that of the IQHE cylinder.

To be concrete, consider a CDW on a finite chain of $N$ sites described by the Hamiltonian:
\begin{align}
{H} =& -t \sum_{j=1}^{N-1}  \left( c^{\dagger}_j c^{\phantom \dagger}_{j+1} + c^{\dagger}_{j+1}c^{\phantom \dagger}_{j} \right) \notag \\  
 &+ V \sum_{j=1}^{N-1} c^{\dagger}_j c^{\phantom \dagger}_{j} c^{\dagger}_{j+1} c^{\phantom \dagger}_{j+1} +\sum_{j=1}^N \zeta^{\phantom \dagger}_j \, c^{\dagger}_j c^{\phantom \dagger}_{j}.
\label{eq:cdw_hamiltonian}
\end{align}
Here, $t>0$ is the tunneling amplitude, $V$ is the strength of the nearest-neighbor Coulomb interaction between electrons, and $\zeta_j$ is a random on-site potential that describes the effect of impurities. The operator $c^{\dagger}_j$ ($c^{\phantom \dagger}_j$) creates (annihilates) a spinless electron at position $x = ja$, where $a$ is the lattice constant. The filling factor $n$ is a common fraction, so that $N n$ is the total number of electrons in the system. As usual, we assume the system to be charge neutral in total, and ignore any ionic charges.

Using the mean-field Ansatz $\langle c^{\dagger}_{j+1}c^{\phantom \dagger}_{j+1}+c^{\dagger}_{j-1}c^{\phantom \dagger}_{j-1} \rangle \approx 2\langle c^{\dagger}_{j}c^{\phantom \dagger}_{j} \rangle = A\cos (Q ja+\phi)$ for the particle density, the interaction term becomes $ \sum_j VA \cos (Q ja+\phi)  c^{\dagger}_{j}c^{\phantom \dagger}_{j}$~\cite{cdw1}. This Ansatz defines the propagation vector $Q=2\pi n / a$ of the charge-density modulations, as well as the phase $\phi$, which determines the position of the CDW with respect to the lattice. Substituting the interaction back into equation~\eqref{eq:cdw_hamiltonian} yields the full mean-field Hamiltonian for the $1D$ CDW chain. For periodic boundary conditions and $\zeta_j=0$, this form of the Hamiltonian will coincide with Hofstadter's tight-binding description of the IQHE~\cite{hofstadter}, if we make the identifications $k\to k_x$, $Q \to e B /(\hbar c)$, and $\phi \to k_y a$. Here, $k$ is the momentum in the CDW, and $k_{x,y}$ are the $x,y$-components of momentum in the IQHE. The radial magnetic field $B$ in the IQHE cylinder determines the filling of its Landau levels. Under this mapping, adding flux along the central axis of the IQHE cylinder corresponds to changing the value of the phase $\phi$ in the CDW~\cite{intro3}, and thus to sliding the charge-density wave along the chain.

The Hamiltonian itself may be realised as a mean-field description of charge order in $1D$ chains of aligned orbitals within a three-dimensional material~\cite{Gruner}, or as an effective description of cold atoms in an optical lattice~\cite{cdw1,Ha2015rme}. Here, we focus first on the topological properties of the theoretical mean-field model, before discussing possible experimental probes of the various emerging modes of topological transport in more realistic settings. 
%
\begin{figure}[t]
\includegraphics[width=\columnwidth]{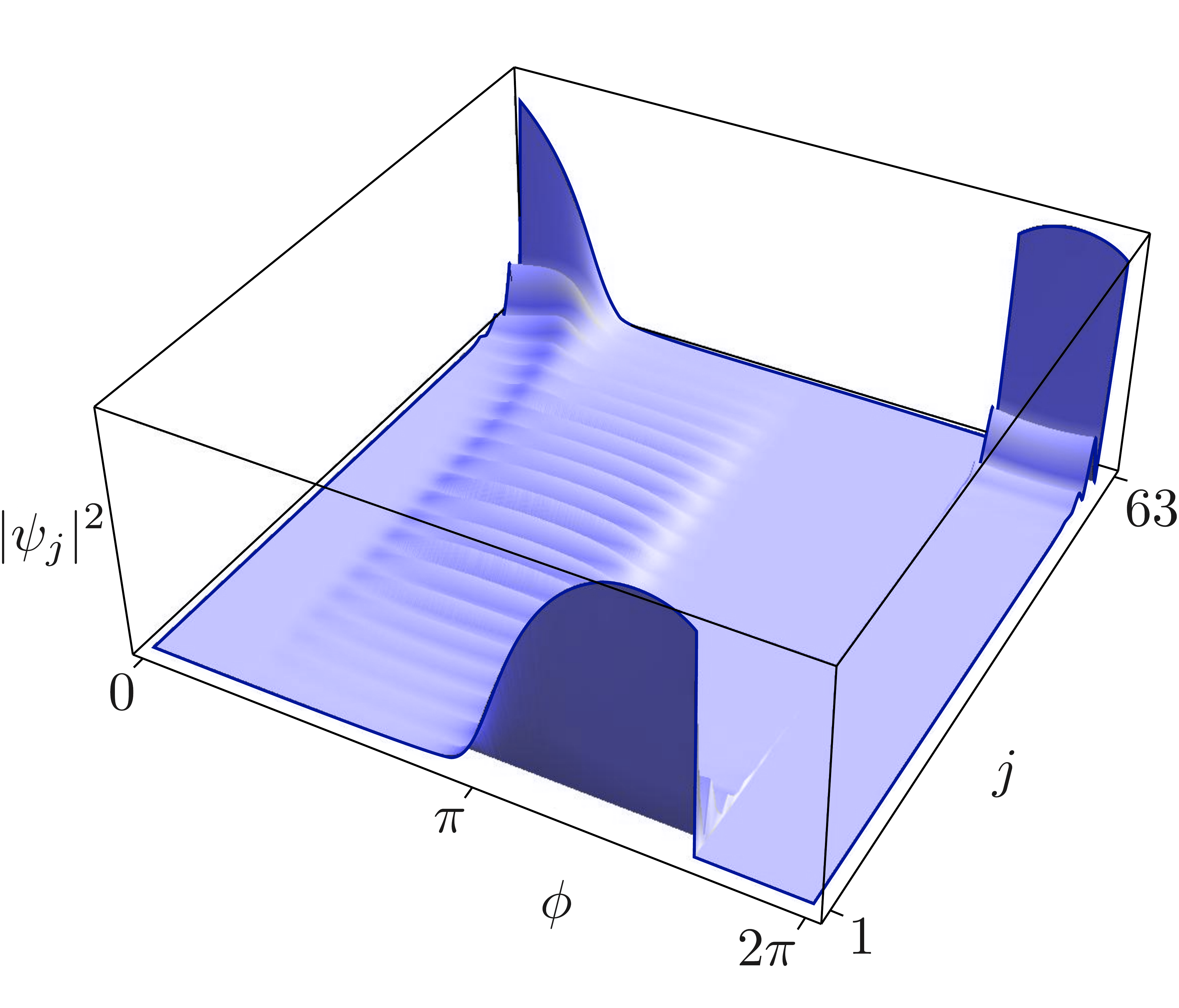}
\caption{\label{fig2} The squared amplitude of the highest occupied wave function at $1/3$ filling, as a function of phase $\phi$ and position $j$ along the CDW chain. The bulk state between $\phi\approx\pi/3$ and $\phi\approx\pi$ can be recognised as a particle in a box eigenfunction with additional CDW modulations. It undergoes an avoided crossing at $\phi\approx\pi$ and becomes a left edge state for values of $\phi$ up to $5\pi/3$, where it changes abruptly into a right edge state until coming back to $\phi\approx\pi/3$. If opposing edges of the chain are connected, the abrupt tunneling from left to right will turn into an avoided crossing of its own, whose adiabatic traversal may correspond for example to the physical process of charge being transferred from one edge to another through an intermediary wire.}
\end{figure}

\section{Visualising charge transfer between edge states}
In Laughlin's argument for quantised transport across the IQHE cylinder~\cite{intro3}, the electromagnetic gauge structure plays a central role. Although helpful in establishing why the charge transferred between the ends of the cylinder must be integer, the presence of gauge freedom makes it hard to directly visualize the precise dynamics. We know and understand what happens upon insertion of a single flux quantum, but questions like at which value of the flux one edge state becomes unoccupied, and the opposite one occupied, or what the wave function looks like after insertion of only half a flux quantum, are difficult to answer in a gauge-independent fashion. For the CDW system, this problem does not exist. The process corresponding to a flux insertion is the sliding of the CDW by precisely one wave length, and this results in a precisely quantised amount of charge being transferred from one end of the chain to the other (Thouless pumping~\cite{thouless}). The quantised conductivity is determined by the sum of Chern numbers for all  occupied bands (in $k,\phi$-space), as in the IQHE. However, in this case we can plot the charge distribution for both bulk and edge states for any value of $\phi$, and hence visualize the topological transport as a continuous process.
%
\begin{figure*}[t]
\includegraphics[width=\textwidth]{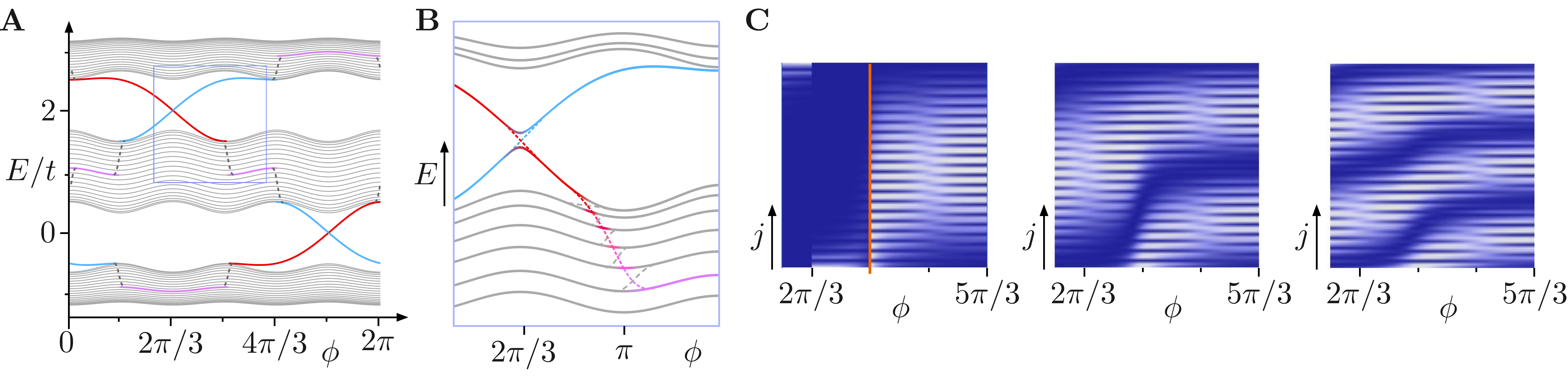}
\caption{\label{fig3} {\bf A} Sketch of the spectrum as a function of the phase $\phi$ with all states related to the topological order indicated. Red lines crossing bulk gaps are edge states localised at the left ($j=1$) end of the chain, blue lines are right ($j=63$) edge states, and purple lines in the middle of bulk bands represent mobility edges. {\bf B} Schematic close-up view of the crossing of two edge states, and of an edge state entering the bulk band. If the two edges are connected they undergo an avoided crossing, which can be either adiabatically traversed by tuning $\phi$ very slowly, or non-adiabatically, by increasing $\phi$ more rapidly in the region of the avoided crossing. As the edge state enters the bulk band, it necessarily undergoes a series of avoided crossings, which ultimately connect the edge state to the mobility edge. {\bf C} Intensity plot of selected wave function amplitudes $|\psi_j(\phi)|^2$, for different values of the phase $\phi$. In all plots, dark blue and white indicate low and high amplitudes respectively. The leftmost plot shows the state which in panel {\bf B} starts out as a right edge state (blue line) for low $\phi$, then becomes a left edge state (red line), and after the avoided crossing at $\phi\simeq\pi$ becomes a bulk state (grey line). The orange line in panel {\bf C} indicates that separate color scales were used for the edge and bulk states. The other two panels show the states that start out at low $\phi$ as the topmost state in the lower bulk band and the one below it. The avoided crossings at $\phi\simeq\pi$ can be identified by  the changing number of nodes in the wave function.}
\end{figure*}

As shown in figure~\ref{fig1}, the spectrum and eigenstates can be computed for a specific choice of parameter values, including any value for the phase $\phi$. As an example, we show the results for a period-$3$ CDW with $n=1/3$, $N=63$, and $AV/t=5$. We use this specific example throughout the paper, but note that taking other parameter choices does not qualitatively affect any of our results. Among the numerically obtained wave functions, bulk and edge states can be easily distinguished. The bulk wave functions for a CDW with open boundary conditions can be understood as products of plane waves (solutions of a CDW system with periodic boundary conditions), and the eigenstates of a particle in a box. The edge states, on the other hand, are exponentially localized on one side of the chain. 

Upon varying $\phi$, the highest occupied state changes from an edge state localized on one side to that on the opposite side. In the example of $1/3$ filling shown in figure~\ref{fig2}, the highest state with $E<0$ at $\phi=0$ is an edge state localised at the right side of the chain ($j=63$). In the region $\pi/3 \lesssim \phi\lesssim \pi$, the edge state is adiabatically transformed into a bulk state, until it emerges again as an edge state on the opposite side of the system. At $\phi \simeq 5\pi/3$, the highest occupied state discontinuously changes from being localised on the left side of the chain ($j=1$), to being localised on the right. From the spectrum, it is clear that this behaviour stems from the two edge states crossing in energy at this point. 

Notice that the states displayed in figure~\ref{fig2} are the highest occupied states at zero temperature for a given value of the phase variable $\phi$. Starting from a phase value $\phi < 5\pi/3$ and adiabatically sliding the CDW forward, the system will in fact not stay in the instantaneous ground state. The two edge states crossing within the bulk gap are located at opposite edges of the chain in real space, and any matrix element of local operators that could assist in tunnelling across is exponentially small in the chain length. A state on one end of the system can therefore not simply jump to the other end in the way suggested by figure~\ref{fig2}. For a sufficiently long CDW, adiabatic variation of $\phi$ causes the system to end up in an excited state at $\phi=2\pi$, with a high-energy edge state occupied and a lower-energy edge state empty. The system can only return to its instantaneous ground state, and the topological material can only function as an adiabatic charge pump, if the two edges of the CDW chain are connected to one another through some external coupling. In fact, in any experimental implementation of a topological quantum pump, one would indeed include a wire connecting the two sides of the system, and typically measure the current through the wire as the phase is being varied. Such a connection can be easily included in the simulation as a weak hopping element between the opposing sides of the chain. It allows the two edge states to interact, and turns their intersection in the spectrum into an avoided crossing by opening up a small energy gap. Adiabatic evolution, which retains the instantaneous ground state throughout a pumping cycle, is once again possible. Varying $\phi$ from $0$ to $2\pi$ then results in the transfer of precisely one (the Chern number of the lowest band in $k,\phi$-space~\cite{cdw1}) electron from one side of the system to the other, through the connecting wire, and thus realises the Thouless pump, or in the IQHE interpretation of the same system, Laughlin's gauge argument. 

\section{The connection of edge states to the mobility edge}
Inspecting the bulk wave functions in the spectrum of figure~\ref{fig1}, two features stand out. First of all, as the edge state enters the bulk band, it does not simply disappear. Tracking for example the highest occupied state in figure~\ref{fig1}, it is clear that the state, which was an edge state at $\phi<\pi/3$, becomes a bulk state at $\pi/3 \lesssim \phi\lesssim \pi$. The bulk state is the particle in a box state with the highest available number of nodes, dressed with charge-density modulations. A state with the same number of nodes in fact already existed for $\phi<\pi/3$, as part of the occupied bulk states. The edge state therefore does not evolve into a new bulk state as $\phi$ is adiabatically changed, but rather has an avoided crossing with an existing bulk state and takes over its character. The existing bulk state takes over the character of the edge state and is pushed down in energy in the process. It then has an avoided crossing with the next bulk state, and so on. This pattern is shown schematically in figure~\ref{fig3}.

\begin{figure}[t]
\includegraphics[width=\columnwidth]{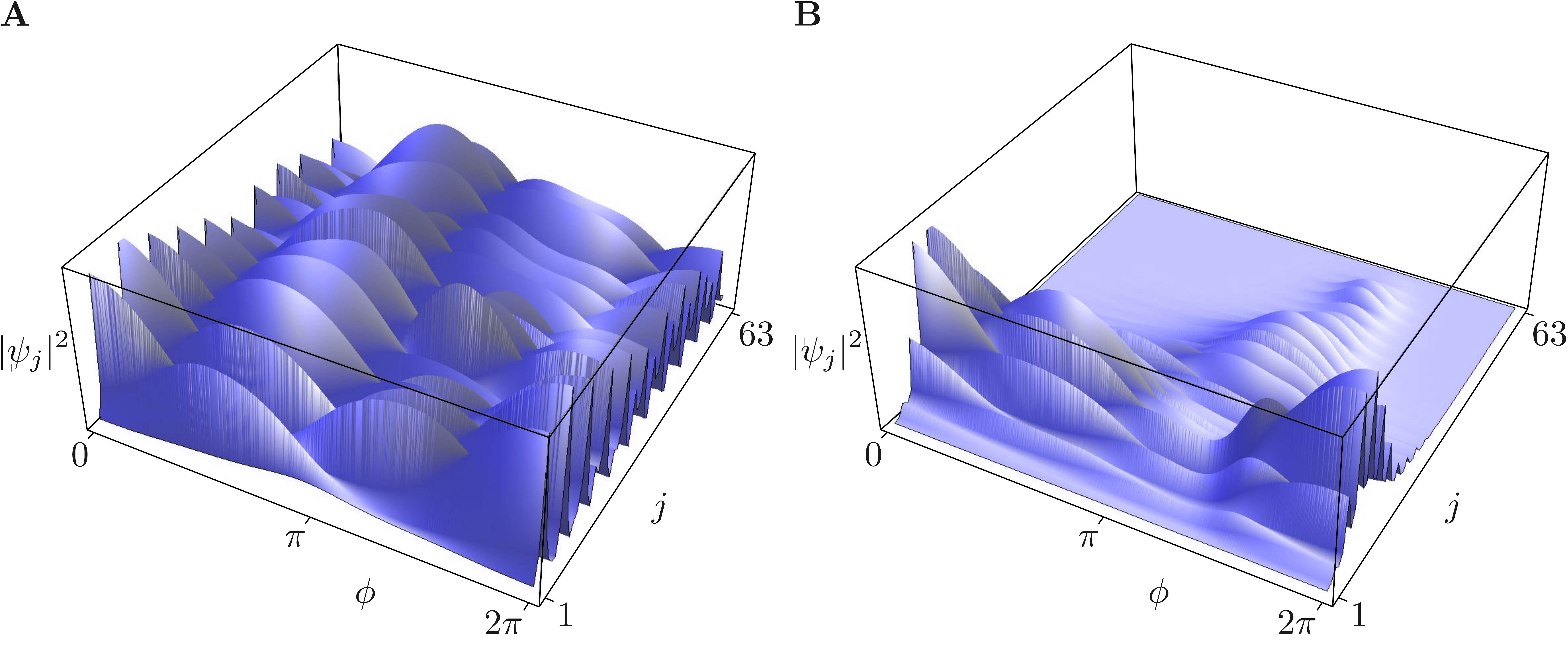}
\caption{\label{fig4} Wave functions in the presence of randomly distributed weak impurities ($|\zeta_j|/t<0.2$). The delocalised mobility edge in panel {\bf A} is hardly affected by the impurities, testifying its topological character. The lowest energy state in panel {\bf B}, on the other hand, becomes completely localised at the location of the strongest impurity. For both panels, the same distribution of impurities was used.}
\end{figure}

Close inspection of the wave functions in the spectrum indeed shows a cascade of avoided crossings, witnessed by a remainder of an exponential localisation that is visible in bulk states undergoing the avoided crossings. These continue until the edge state reaches precisely the middle of the bulk band. There, it emerges again for an extended range of $\phi$, in the form of another well-known topologically special state, the mobility edge~\cite{RMP_me}. This isolated critical state was to be expected in the centre of the bulk band, because the extended bulk states generic to disordered system are known to be constricted to just a single energy precisely in the centre of each (impurity-broadened) Landau level in the case of the quantum Hall effect~\cite{Chalker_me}.  Just like the mobility edge was formed from a left edge state at $\phi\simeq\pi/3$, it goes through a second series of avoided crossing and reemerges as a right edge state at $\phi\simeq\pi$. The edge states are of the form $\psi(x) \propto e^{\lambda x}$, with $\lambda$ negative or postive for edge states localised at small or large $x$ respectively. The mobility edge naturally connects these two states, and has the same shape of wave function, with $\lambda=0$. This special, totally delocalised wave function is thus a plane wave connecting both sides of the sample. Within the CDW, the plane wave is again dressed by charge modulations, as shown in figure~\ref{fig4}. Note that for even $Nn$, it is possible to have two orthogonal, delocalized wave functions. We consider chains with odd $Nn$ here, so that a single mobility edge connects to edge states in both bulk gaps. 
%

The topological nature of the mobility edge becomes apparent upon adding a random impurity potential $\zeta_j$. Notice that in contrast to the IQHE, where the presence of weak impurities is necessary to observe the quantisation of the Hall conductivity, in the CDW system topological transport may be observed even with only the Coulomb interaction localising charges within the unit cell. Nevertheless, the presence of weak impurities (modeled here with $|\zeta_j|/t<0.2$) may be considered, and their effect is much the same as in the IQHE, localising electronic states at specific locations in the chain. The difference between the mobility edge and other bulk states is now immediately obvious. As shown in figure~\ref{fig4}, typical bulk states are severely affected by the impurity potential, being amplified and suppressed at random locations. The mobility edge on the other hand, retains a more or less constant amplitude along the entire length of the chain, as long as the impurity strength is weak. This clearly indicates the topological nature of the mobility edge, and hence its connection with the topological edge states.

\section{Non-adiabatic quantised particle transport}
The connection between edge states and mobility edge may be illustrated by considering a non-adiabatic mode of quantised transport, which exists in addition to the well-known adiabatic Thouless pump. Starting for example with the (red) left edge state just below $\phi\simeq 5\pi/3$ occupied, an induced non-adiabatic evolution that jumps all avoided crossings with both bulk and edge states would result in the highest occupied state going around the full spectrum as shown schematically in figure~\ref{fig3}. As $\phi$ is increased by $6\pi$, the highest occupied state has traversed all topologically special states in the entire spectrum, and returns to its initial configuration. Stopping the non-adiabatic evolution after $\phi$ is increased by $4\pi$, however, the highest occupied state will have moved from its initial left edge state to the right edge state in the lowest bulk gap. In the presence of a wire connecting both ends of the chain, we can then either adiabatically increase $\phi$ by another $2\pi/3$, or let the system spontaneously relax to its ground state. Both will lead to the excited state going from right to left through the wire. That is, a current is carried through the connecting wire in the direction opposite to that of the usual adiabatic topological transport. This non-adiabatic transfer of charge with the same magnitude but in the opposite direction of the usual Thouless pumping, is made possible by the edge states and mobility edges forming a connected set of states winding throughout the electronic spectrum.

Generalising the procedure to systems of any filling, the integer number of charges transferred by the non-adiabatic mode of transport will equal $-C$, with $C$ the total Chern number of all occupied bands, the exact opposite of the $C$ electrons conveyed in the usual Thouless pump. In this sense, it acts as a sort of anti-Thouless pump. Notice that hybrid protocols employing both adiabatic and non-adiabatic driving can accomplish the same thing in a much simpler fashion. In such protocols, however, the highest occupied state is in non-topological, localised bulk states for part of the pumping cycle. The purely non-adiabatic process suggested here, in which the highest occupied state jumps all avoided crossings that it encounters, is special because it depends entirely on the special nature and connectivity of the topological edge states and mobility edge states.

The non-adiabatic topological charge pump presented here is intended purely as an illustration of the connection between different types of topological states. Nevertheless, non-adiabatic or hybrid protocols may in some aspects be preferable to purely adiabatic ones in practical implementations of topological pumping. As has been recently pointed out, the presence of a spectrum of unoccupied states imposes much more stringent conditions for achieving adiabaticity than just the time scale of the driving being longer than the inverse gap size~\cite{cheianov1,cheianov2,gritsev}. Non-adiabatic driving across an avoided crossing on the other hand, can be accomplished by sufficiently fast changes of the driving parameters, as long as the overlap between initial and target states is large. We confirmed that this is the case in the model CDW system by achieving an almost complete transfer of occupation across avoided crossings upon ramping up the driving speed. For any experimental implementation, a practical restriction on parameter values may arise from a preference for maintaining the validity of the mean-field solution even under non-adiabatic driving. 

\section{A possible implementation}
To realise the proposed illustration of a non-adiabatic pumping cycle in experiments, several techniques for imaging the real-space structure of CDW systems can be used. First of all, the static real-space structure of a CDW can be imaged by scanning tunneling microscopy (STM), which directly observes the amplitude and phase of charge-density modulations induced by a CDW~\cite{stm1,stm2,stm4,stm5}, as well as excess charge at the end of a chain~\cite{est_exper}. Straightforward experimental realisations of the dynamic quantised transport protocols, on the other hand, may use ultracold atoms in an optical lattice or photons in a waveguide array. Waveguide arrays may directly simulate the CDW Hamiltonian~\cite{est_exper}. Variations of the mean-field CDW phase $\phi$ are then implemented by appropriate variatons of the index of refraction. The quantised transport of charge can be observed by injecting photons into the waveguide at one edge and observing the intensity distribution in the waveguides after various propagation distances. The required values for experimental parameters are similar to those used in the literature~\cite{est_exper}, and can be realistically obtained with existing technology.

Using ultracold atoms, the mean-field Hamiltonian can be constructed by projecting a circular optical dipole lattice of $63$ sites through a microscope objective~\cite{Ha2015rme,Lohse2018e4q}. This can be done for example using an objective with a numerical aperture of $0.8$~\cite{Atoms1}, and a short wavelength of $532$\,nm~\cite{Atoms2}, obtaining a waist of the projected Gaussian beams of the order of $0.4$\,$\mu$m. To ensure that the projected potential is attractive at this wave length, one could use, for example, Sr atoms to load the lattice with. Gaussian beams with a spacing of $0.7$\,$\mu$m then result in a potential that closely resembles a sine. The lattice depth will be approximately 0.3 times the depth of one Gaussian beam. The lattice sites can be created by imaging a mask or a pattern from a digital mirror device (DMD) through the objective ~\cite{Ha2015rme}. In order to confine the atoms in the direction orthogonal to the plane of the ring, an optical lattice can be applied in that direction.

Appropriate values of the model parameters, establishing a Mott-insulator regime with a weak residual tunneling, may be achieved for example by choosing a lattice depth of about eight recoil energies, or approximately $10$\,kHz. The mean-field energies are then $V = 270$\,Hz and $t = 45$\,Hz. A weak link may be introduced between two particular sites in the lattice to allow them to act as edges. The strength of the tunneling across the weak link can be adjusted arbitrarily far down from $45$\,Hz by increasing the spacing between the two selected sites. 

To create and move the CDW, an additional lattice with larger spacing may be added~\cite{intro13}. The required specifics for this additional lattice are much more relaxed than the ones of the primary lattice, and can be freely adjusted within a sizeable range. The rotation of the secondary lattice could be achieved by imaging a DMD or a rotating mask onto the atoms. The lifetime of the system is limited by off-resonant scattering of the strongest, primary lattice, which for the parameters discussed above, will be over $200$ seconds. The experiment should take a few times 63 tunnel times, which is a few times 2.8 seconds, and therefore comfortably fits in the expected experimental lifetime. The system can be prepared in its ground state by using the Mott-Insulator transition. After a full driving cycle, the quantised build-up of charge at the edges can be detected by quantum gas microscopy, which directly measures the parity of the number of atoms in each lattice well. All of this can be realistically done with existing technology.

\section{Conclusions}
We propose a procedure to visualize adiabatic and non-adiabatic topological charge transport in a $1D$ charge-ordered system, which is well-known to map precisely onto the tight-binding model for the $2D$ IQHE in the Landau gauge used by Laughlin to explain topological transport in an IQHE cylinder. The absence of an electromagnetic field in the CDW, and hence of the need to make a gauge choice, enables us to directly compare plots of wave functions at different values of the CDW phase. The topological transport and dynamics of electrons in the CDW chain found here, thus give direct insight into the detailed motion of electronic states in the IQHE as well.

The visualisation of the topological transport brings to the fore an explicit connection between edge states localised at the ends of the chain or cylinder, and the mobility edges residing in the very centre of their bulk states. To illustrate the connection between these two types of topological states, localised and extended, we formulate a purely non-adiabatic protocol that nonetheless results in quantised, topological transport of charge. The number of electrons pumped in a single cycle is precisely equal to that of the well-known adiabatic pump, but they flow in the opposite direction, creating a sort of ``anti-Thouless'' pump. We expect our conclusions, and in particular the connection between edge states and the bulk mobility edge to apply more generally to disordered topological systems. Any system with a non-trivial Chern or Z$_2$ invariant is guaranteed to have edge states, and similarly mobility edges arise generically in models with disorder-induced localisation. A simplified intuitive picture of the connection between the two types of states, suggested by the CDW system studied here, can then be drawn starting from an edge state localised on one side of the system. By evolving this state as a function of some system parameter, such as the CDW phase, it may be adiabatically connected to an edge state on the opposite side of the system. If the evolution is adiabatic throughout, however, a fully delocalised state must generically exist between the two edge-localised extremes.

Finally, we suggest experiments on ultra cold atoms in an optical lattice, as well as in photonic waveguides, which can test the proposed connection between edge states and mobility edges, as well as the non-adiabatic pumping cycle, using realistic parameter values. 

\subsection{Acknowledgments}
This work is part of the Delta Institute for Theoretical Physics (DITP) consortium, a program of the Netherlands Organization for Scientific Research (NWO) that is funded by the Dutch Ministry of Education, Culture and Science (OCW). JvW acknowledges support from a VIDI grant financed by the Netherlands Organization for Scientific Research (NWO). This project has received funding from the European Research Council (ERC) under the European Union's Seventh Framework Programme (FP7/2007-2013) (Grant agreement No. 615117 QuantStro).

\bibliographystyle{apsrev4-1}
\bibliography{topotrans}
\end{document}